\documentclass[12pt]{revtex4}

\usepackage{psfig,epsfig,array}

\begin{document}

\title{Tailoring the photonic bandgap of porous silicon dielectric mirror}
\author{V. Agarwal* and J.A. del R\'{\i}o }
\affiliation{Centro de Investigaci\'on en Energ\'{\i}a\\
Universidad Nacional Aut\'onoma de M\'exico\\
AP 34, Temixco, Mor. CP 62580, M\'{e}xico.\\
* present address CIICAp-Universidad Aut\'onoma del
Estado de Morelos, 
Av. Universidad 1001, Col Chamilpa, Cuernavaca, Mor, M\'exico}

\begin{abstract}

A systematic method to fabricate porous silicon one dimensional photonic 
crystals has been
engineered to have a photonic bandwidth up to 2000nm. The observation of the 
tailorability of the photonic bandgap (PBG) underscores the requirement of the 
large refractive index contrast for making broad PBG structures. In this
letter, we present the fabrication and characteristics of 
such structures that may be promising structures for a large variety of
applications.
\end{abstract}

\maketitle

\newpage

Dielectric mirrors can reflect light for any direction of propagation 
in specific wavelength regions\cite{john}- \cite{john1}. 
This property will allow the creation of all-optical integrated circuits 
as it can be used to confine, manipulate and guide the photons.
The advantage of a dielectric mirror over metallic ones
is due to the dispersive and absorbing regions in the visible and infrared
spectra of metals \cite{handbook}. 
This kind of dielectric photonic band gap (PBG) structures have already been 
used in vertical-cavity 
surface-emitting lasers \cite{breiland}, dielectric interference 
filters \cite{Hetch}, sensors \cite{sensor} and other devices.

Alternating layers of material with different dielectric constant are the 
simplest possible photonic crystals known so far. 
These dielectric 
multilayer films act as a perfect mirror for light with a frequency 
within a sharply-defined gap. For a fixed number of periods, this PBG 
increases, as the ratio of the refractive indices of the two layers
($n_H/ n_L$, high, H, and low, L, refractive index $n$) increases 
\cite{john},\cite{Pavesi}, 
and for a fixed ratio of refractive 
indices, upto a certain limit, the PBG increases with the number 
of periods \cite{Pavesi}. In this letter we propose an idea to tailor the PBG 
structures, ranging from few nanometers 
to 2000 nm or more, for two different values of refractive indices,
($n_H,n_L$) without the prerequisite of a huge contrast.  The idea 
has been  successfully implemented experimentally on porous silicon.

Porous Silicon (pSi) is already considered as a promising material
for photonic applications \cite{selena}. This material 
can be fabricated with electrochemical etching of silicon with HF. In 
this material, the porosity is a
linear function of the current density for a specific HF concentration and 
anodization time \cite{Pavesi} and 
is known to show a refractive index contrast for different porosites
\cite{theiss}.  
Therefore, a periodic pulse 
alternating between two different current densities has been a convinient 
procedure to fabricate multilayer pSi films \cite{Pavesi},\cite{selena},
\cite{Canham}. 
The change of
the  current density does not affect the previously formed pSi layer because 
silicon dissolution occurs at the silicon electrolyte interface \cite{Theib}. 
 
The multilayer pSi one dimensional-PBG 
structures known till now had the PBG of less than 500nm 
\cite{Pavesi} \cite{selena}. 
What we require is an easy method to modify the PBG of the 
structure according to our requirements. 
For the first time, through this letter, a technique to tailor the  
PBG structures, is successfully materialized on pSi multilayer 
mirrors.
Stacks of $\lambda /4 $ ($n_H d_H = n_L d_L = \lambda /4 $, with $d_i $ 
being the real thickness) mirrors have been prepared for different wavelengths 
in the visible and near infrared region (one submirror after the other 
submirror). 
The number of submirrors fabricated in a single structure varies from 2 to 80.
Each mirror is designed 
for a different wavelength and can consist of 2.5 to 20 periods according 
to the requirements and the region we want to work with. Thus, the number of 
pSi layers can go up to 700. The exact selection of the $\lambda $-values 
depends on the application which will be discussed in the latter part of the 
paper.

Our method has been inspired by the ladder structure
simulated by Kavokin et al. \cite{kavokin} in order to observe photonic 
Bloch oscilations. The structures proposed are coupled
microcavities, but here we present the structures without cavities, 
considering only dielectric Bragg mirrors (DBM). 
Of course, ladder structures can be
prepared by different techniques, i.e., Molecular Beam Epitaxy, 
Metal Oxide chemical Vapor Deposition, etc. 
However in this 
letter, we use a cheap, easy and fast techinque to produce the same.

We have fabricated pSi multilayers by wet electrochemical
etching \cite{viv} of highly boron doped substrate (p+, 0.001-0.005 
$\Omega cm$).
In order to have better interfaces, an aqueous HF/ethanol/glycerol 
electrolyte with 15\% HF, 75\% ethanol and 10\% glycerol \cite{servidori}
concentration is used to anodize the silicon substrate. In addition, 
in order to maintain a constant HF concentration over the interface between
Si and pSi under chemical attack, 
during the etching process
a peristatical pump is used to circulate the electrolyte within the Teflon 
cell. Anodization begins when a constant current
is applied between the silicon wafer and the electrolyte by means of an 
electronic circuit controlling the anodization process.
To stabilize the pSi multilayer structure, all the samples are 
thermally oxidized in an oxygen ambient at 900 $ ^0$C for 10 min 
\cite{fauchet}.
 Oxidation 
induces a blue shift in the peak reflectivity due to the decrease in the 
refractive indices of the layers.  
In our structures, the multilayers have alternating  porosities of the order 
of
$ 35-55\%  (n_H) $ and $ 60-85\% (n_L)$. The time taken in the etching 
process varies from few minutes to 3 hours.
Hence to make the surface stable, the first layer is always a low porosity 
layer and made 10-20 nm thicker than usual for thick structures (more than 
20 microns). 
The refractive indices of the pSi layers have been estimated using 
reflectivity spectra of 2 $ \mu $m thick single layers at 1500nm.
Scanning electron Microscopy (SEM) was used to 
examine the structural features of the films.
The reflectivity spectra of the samples are taken  
with a Shimadzu UV3010 ultraviolet-visible-near infrared spectrophotometer 
at 5 degree incidence.

\begin{figure}[h]
\begin{center}
\resizebox{!}{10.5cm}{\includegraphics{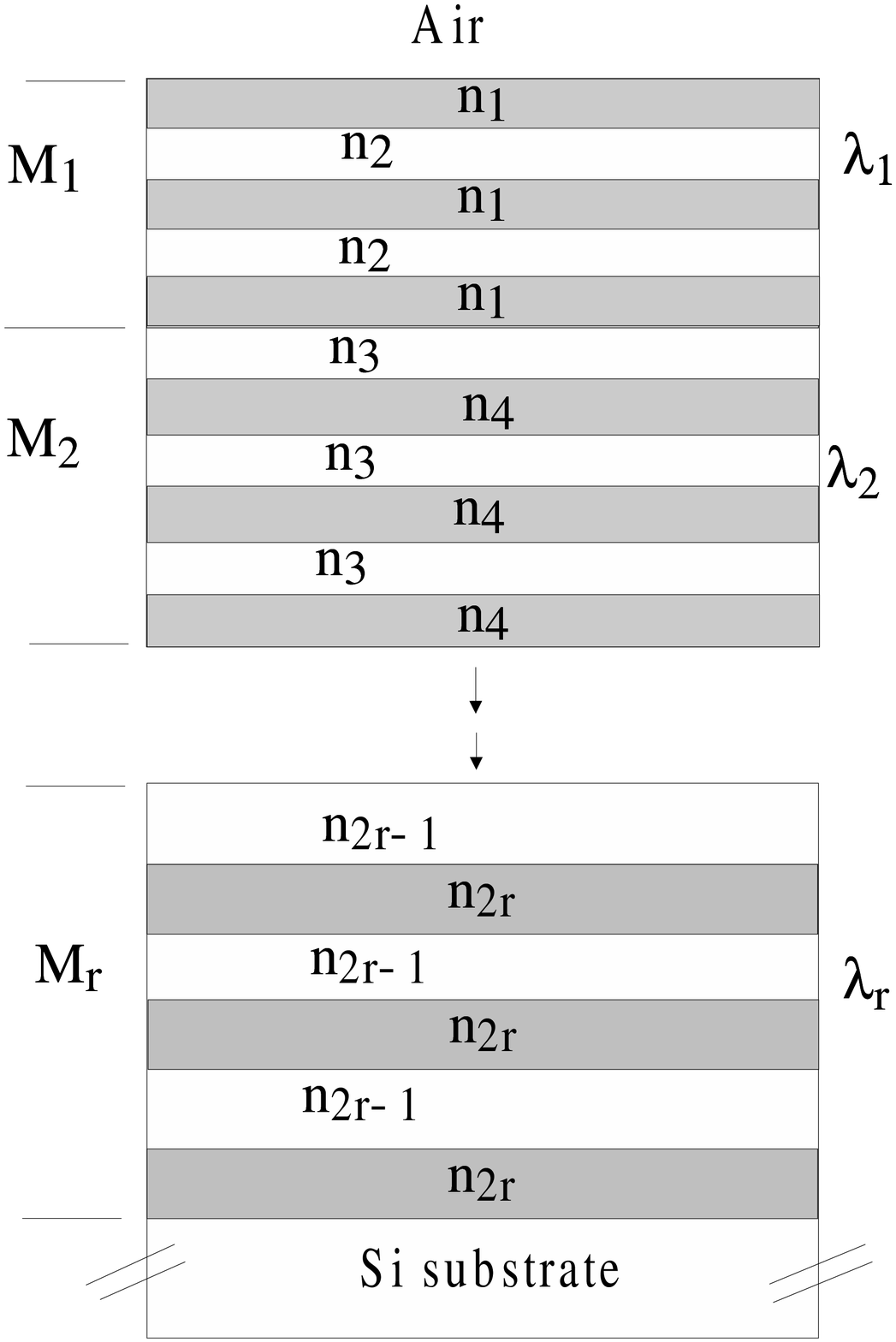}}
\end{center}
\caption{Figure 1. Schematic draw of the proposed structure. The arrows
indicate the direction of the etching process, $i={1, 2,...r} $ indicates
the number of $\lambda /4 $ submirrors in the complete structure.
The number of
submirrors depends on the desired range of reflection.  The number of
layers present
in each submirror affect the flattness of the photonic band gap. $n_i $
represents
the refractive index of each layer. The refractive index contrast
$n_{2i} / n_{2i-1} $, can be different/same for different submirrors.}
\label{schema}
\end{figure}

The schematic of the structures can be seen in Fig.\ref{schema}. The 
idea is to make one submirror after another submirror like a continuum
array of submirrors. The specific range of $\lambda $-values depends directly
on the application. For instance,  a mirror 
reflecting the complete near infrared (NIR) range (800 to 2500 nm) only; 
for this we need to select
the different wavelengths in this specific range in a way that the 
PBGs of respective submirrors overlap and 
give us a continuous PBG reflecting the full NIR range. For designing the 
mirrors 
for the latter part of the visible region and the NIR region, we have 
empirically selected the $\lambda $ values as follows. First select the 
starting $\lambda_1 $-value and then the subsequent $\lambda $-values obey 
a simple relation 
$\lambda_{i+1} - \lambda_i = 2+i $, where $i $ represents the submirror 
number. This selection has been performed and shown in the latter part of 
the letter.

We have made various structures composed of single mirror with 10 periods
(one period consists of one low porosity and one high porosity layer)  
up to a mirror composed of a set of 80 submirrors. 
The total number of layers varies from  20 to 650 respectively. Taking into
account that in the wide PBG structures the anodization time is of the
order of hours,
 to make the structure more stable the first mirror 
always starts with a high refractive index layer, consisting of 2.5 to 4.5 
periods.
The subsequent submirrors consists of 3 to 5 periods reducing the
etching time without affecting the flatness of the photonic band gap.

\begin{figure}[h]
\begin{center}
\resizebox{!}{8.5cm}{\includegraphics{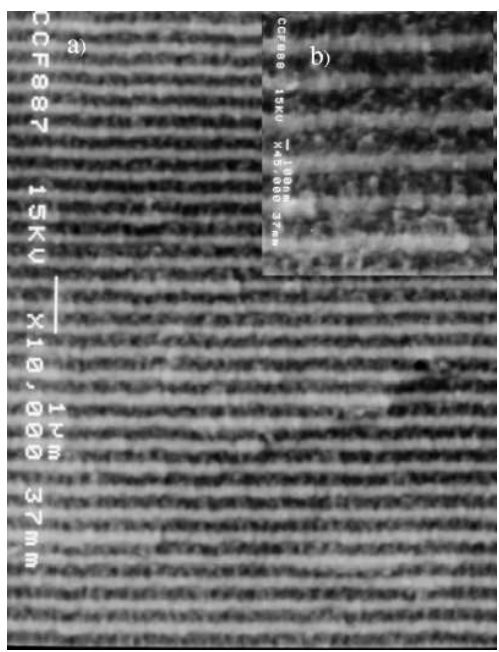}}
\end{center}
\caption{Figure 2. Cross sectional SEM image of a mirror designed for partial
visible and complete near infrared region. Inset shows a amplified
 view of the structure.}
\label{sem}
\end{figure}

\begin{figure}[h]
\begin{center}
\resizebox{!}{8.5cm}{\includegraphics{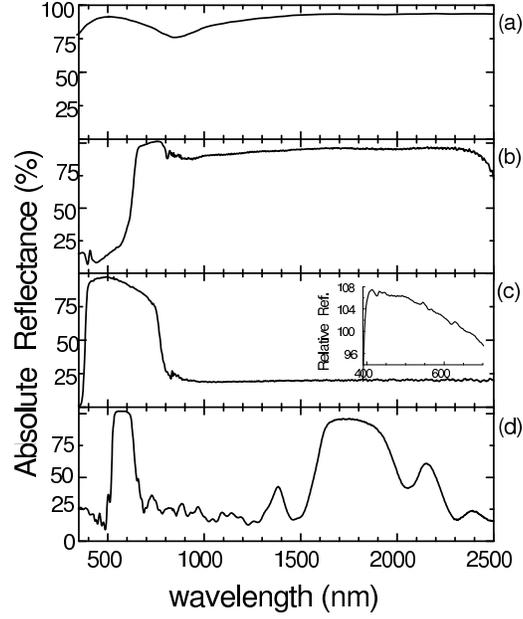}}
\end{center}
\caption{Figure 3. Absolute reflectance vs wavelength for 
(a) Commercially available Al
mirror with SiO overcoat. (b) Partial visible and complete near infrared
region mirror made of porous silcon multilayers (refractive index contrast:
1.4/2.2) (c) complete visble range
mirror made with refractive index contrast 1.4/2.2. Inset showing the
relative reflectance of this mirror with respect to Al (d) a discrete
mirror designed
for 600 and 1700 made with refractive index contrast 1.4/2.2 and 1.5/2.0
respectively}
\label{abs}
\end{figure}

Fig. \ref{sem}(a) shows cross sectional SEM images  of 9 out 
of the 54 submirrors in the complete structure designed for
the partial visible and complete near infrared region. This image 
shows a continuum array of submirrors prepared using 
$\lambda_{i+1} - \lambda_i = 2+i $, designed by taking $\lambda_1 = 700 $
 with a refractive index contrast given by 
$n_L / n_H = 1.4/2.2 $. 
The reflectance spectrum of this structure is shown in Fig.\ref{abs}(b).

The absolute reflectance spectra of some of the structures have 
been shown in Fig. \ref{abs}. The absolute reflectivity of commercially 
available 
Aluminium (Al) mirrors with a SiO overcoat (as bare silicon is prone to 
oxidation and loses its reflectivity) is shown in 
Fig. \ref{abs}(a) \cite{oriel}. The figure clearly shows an absorption in the 
visible region. From 400nm to 2500 nm the reflectivity varies between 85-90\% 
with a dip to 75\%  at around 825nm. Taking this mirror as a reference 
(available with the Shimadzu UV3101 UV-vis-NIR spectrophotometer) the
 reflectivity spectra of our
structures have been measured. Hence in the visible region, the reflectance 
shown 
by most of our structures was higher than that for the Al mirror. 
Thus, the data 
have been corrected for the Al mirror response.
Fig.3(b)-(d) show the absolute reflectance of few examples demonstrating 
the successful implementation of the idea.  In 
fig \ref{abs}(b) we present the response of the mirror shown in 
Fig.\ref{sem}. It has the PBG 
for the partial visible and complete NIR region i.e. from 615nm to 
2450nm. The absolute reflectance can be clearly seen within 85-100\% which 
is clearly 2-20 \% more than that for the Al mirror from 615-1200nm.   
In fig \ref{abs}(c) we plot the absolute reflectance of a mirror 
designed for the complete visible range. The absolute reflectance of this
mirror is higher than the Al mirror from 390 to 650 nm range, as can be
seen, from the inset which shows the raw data coming from Shimadzu UV3101 
UV-vis-NIR spectrophotometer having Al mirror as reference. 
This mirror is composed by 77 submirrors, with the starting 
wavelength
450nm and the subsequent DBMs are determined by the relation 
$\lambda_{i+1} - \lambda_i =3$   
having 4 periods in each submirror. The refractive index contrast used was 
$1.4/2.2 $.

The reflectivity of a mirror reflecting two different wavelength ranges  
has been shown in 
Fig. \ref{abs}(d).
This particular mirror is composed by two submirrors, one prepared for
600 nm ($n_L /n_H = 1.3/2.3 $) and another for 1700 nm ($n_L /n_H = 1.6/2 $)
with 10 periods each. The idea is to show how we can change the refractive 
index contrast from one submirror to another submirror if required.
Theoretical simulations have been
done for all the tailored PBG structures and  seen to be in 
agreement with the experimental reflectivity spectra \cite{va}

Here we can clearly see the enhancement of the PBG width with the 
increase in the number of submirrors. 
We can very well understand that this broad PBG
is due to the overlapping of all the PBGs of the submirrors 
designed for different wavelengths. The advantage of these mirrors 
over metallic mirrors (Al or silver) is their selective nature 
and higher reflectivity. 
For example, designing a mirror for the complete NIR range without
significant reflection from the visible range or viceversa will be excellent 
for different applications.

It is important to stress that Pavesi group too has developed photonic 
structures
with wide photonic band gap using aperiodic porous silicon layers \cite{oton}.

Apart from the structures shown in Fig. \ref{abs} we have tailored the 
PBG for many other mirrors, for example in the spectral ranges of 
356-650nm, 400-600nm, 600-1200nm, 600-2000nm, etc.
The idea is to design a PBG of desired width and 
at required wavelengths. 
Here we present just few examples with a lot of scope for improvement.
It will also stimulate the interest of making better filters 
based on single microcavity \cite{microcavities}, because with the method 
presented here we are able to increase the high reflecting range on both
sides of the cavity.

For the first time, it has been made possible to tailor the PBG 
of the dielectric mirrors, for the partial UV and complete vis-NIR range 
(from 356nm to 2500nm) to have desired width at required wavelenths.
These mirrors are easy to prepare, selective and  more reflective 
than the commercially available Al mirrors.

We gratefully acknowledge the useful discussions with Dr. Miguel Robles 
and help provided by Jose Campos, Dr. Oswaldo Flores (CCF-UNAM).

\newpage

\newpage

\newpage

\newpage

\end{document}